\documentclass[letterpaper]{article}
\usepackage{times}
\usepackage{helvet}
\usepackage{courier}
\usepackage{graphicx}
\begin{document}
%
\title{Combinatorial Analysis of Multiple Networks}
\author{Matteo Magnani \and Barbora Micenkov\'a \and Luca Rossi
\thanks{M.~Magnani is with ISTI, CNR, Pisa (matteo.magnani@isti.cnr.it) and has done part of this work while at the Department of Computer Science, Aarhus University, Denmark, Barbora Micenkov\'a is at the Department of Computer Science, Aarhus University, Denmark (barbora@cs.au.dk), and L.~Rossi is at the Department of Communication Studies, University of Urbino, Italy (luca.rossi@uniurb.it). This work has been supported in part by the Italian Ministry of Education, Universities and Research PRIN project \emph{Relazioni sociali ed identit\`a in Rete: vissuti e narrazioni degli italiani nei siti di social network} and FIRB project RBFR107725.}}

\date{}

\maketitle
\begin{abstract}
The study of complex networks has been historically based on simple graph data models representing relationships between individuals. However, often reality cannot be accurately captured by a flat graph model. This has led to the development of multi-layer networks. These models have the potential of becoming the reference tools in network data analysis, but require the parallel development of specific analysis methods explicitly exploiting the information hidden in-between the layers and the availability of a critical mass of reference data to experiment with the tools and investigate the real-world organization of these complex systems. In this work we introduce a real-world layered network combining different kinds of online and offline relationships, and present an innovative methodology and related analysis tools suggesting the existence of hidden motifs traversing and correlating different representation layers. We also introduce a notion of betweenness centrality for multiple networks. While some preliminary experimental evidence is reported, our hypotheses are still largely unverified, and in our opinion this calls for the availability of new analysis methods but also new reference multi-layer social network data.
\end{abstract}


\section{Introduction}

Recently, large user-generated network data have become available through online social network sites like Twitter \cite{Huberman2009}, Facebook \cite{Stefanone2011}, YouTube \cite{Cheng2008}, Friendfeed \cite{MagnaniSBP2010} and several others. This phenomenon has determined a new wave of interest in Social Network Analysis (SNA), especially when applied to very large networks.
However, while traditional SNA has mainly focused on the analysis of a single platform at a time, recent researches have stressed how both online and offline social experiences can hardly be simplified using a single kind of relationship between homogeneous subjects \cite{Cai2005,Mucha2010,Szell2010,MagnaniASONAM2011,Berlingerio2011,Brodka2011}.
Just like our offline experience is defined as a set of relationships having a specific meaning within a specific context \cite{Goffman1974}, our online experience cannot be reduced to a single flat layer of connections. By contrast, it can be seen as a composite, stratified phenomenon. As it has been pointed out \cite{Pepe11}, we are what the Italian novelist Luigi Pirandello defined as  \emph{one, no one and one hundred thousand} at the same time.

\begin{figure}[t]
\begin{center}
\includegraphics[width=.6\columnwidth]{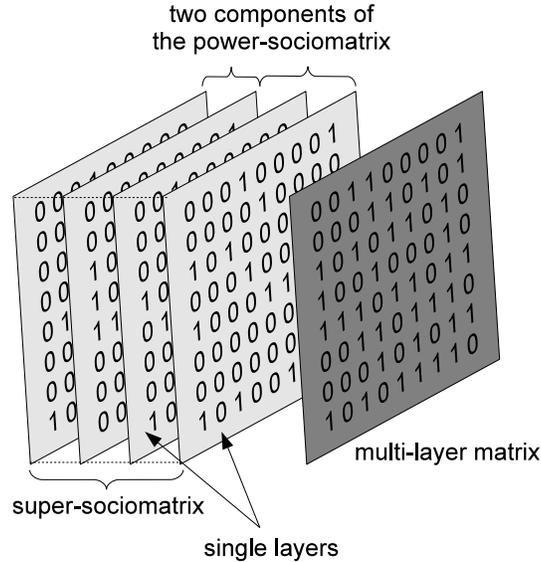}
\end{center}
\caption{Different data structures to study multiple networks: super-sociomatrix, single layers, and multi-layer matrix. The \emph{power-sociomatrix}, that we introduce for the first time in this paper, is made of all possible combinations of layers---mathematically, it is the power-set of the super-sociomatrix}
\label{f:matrices}
\end{figure}

Several models have been proposed to represent multiple related networks (a \emph{multi-layer} network). \cite{Wasserman1994} define the \emph{sociomatrix} of a relation $\mathcal{R}$ as a matrix with entries $x_{ij}$ representing the value of the tie from node $i$ to node $j$ on relation $\mathcal{R}$---in this paper we focus on the case where $x_{ij} \in \{0, 1\}$, i.e., a tie is either present or not. A \emph{super-sociomatrix} is a collection of sociomatrices where each of them corresponds to one type of relation in a multi-layer network. In Figure~\ref{f:matrices} we have represented this data structure: four layers, each one defining a network (e.g., Twitter, Facebook, FourSquare and LinkedIn connections) and composing the so-called super-sociomatrix.

\begin{figure}
\centering
\includegraphics[width=0.4\columnwidth]{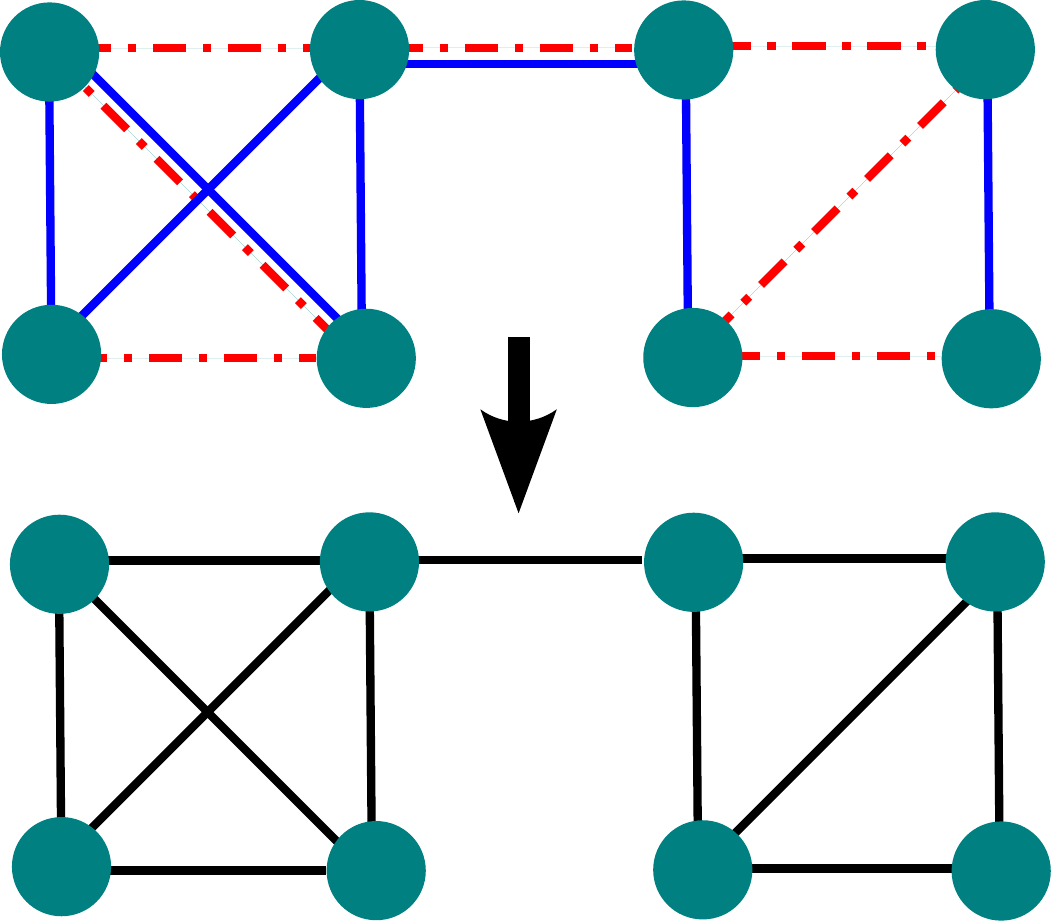}
\caption{A network visualization of a super-sociomatrix: two different layers (kinds of relationships) and a representation of the multi-layer matrix obtained merging these two layers}
\label{fig_defSociomatrix}
\end{figure}

While the super-sociomatrix model has been available for a long time, collecting meaningful multi-layer data is a complex activity, and there is only a limited choice of tools for a specialized analysis of multi-layer networks.
In fact, with a few exceptions \cite{Szell2010,Berlingerio2011,DBLP:conf/kdd/BonchiGGU12,MagnaniSBP13-1}, once these networks have been constructed, they are analyzed using traditional methods, e.g., the different layers are merged and existing network centrality measures are applied. For example, Figure~\ref{fig_defSociomatrix} shows how two different layers (top) correspond to a merged network (bottom) that can be analyzed using any existing SNA tool.

In this paper we propose a methodological shift in the way in which multi-layer networks can be analyzed. This is then instantiated into three main research hypotheses to be tested on a real dataset that we have collected to this aim. The \emph{fil rouge} behind these hypotheses is that the identification of different semantic layers connecting a set of individuals and the analysis of the exponential number of combinations of these layers can allow the identification of hidden patterns. A strong underlying conjecture that we make is that not only it is important to consider multiple layers, as it has been suggested in several related works, but also that considering all the single layers one by one and/or the complete multi-layer network containing all connections may result in information loss. Therefore, we introduce the new concept of \textbf{power-sociomatrix}. Going back to Figure~\ref{f:matrices}, we can see how the power-sociomatrix is made of all combinations of layers---in the figure, two of the 15 elements of the power-sociomatrix are indicated, correponding to the first two and the last three layers respectively. 

Our general hypothesis, if verified, would have an impact on several aspects of network analysis, including the definition of distance between nodes and methods to find communities. However, as we will see in the following, so far we have been able to observe only some of its foreseen effects, leading to the necessity of additional experiments on new datasets.

\subsection{Research hypotheses}

\begin{figure}
\centering
\includegraphics[width=\columnwidth]{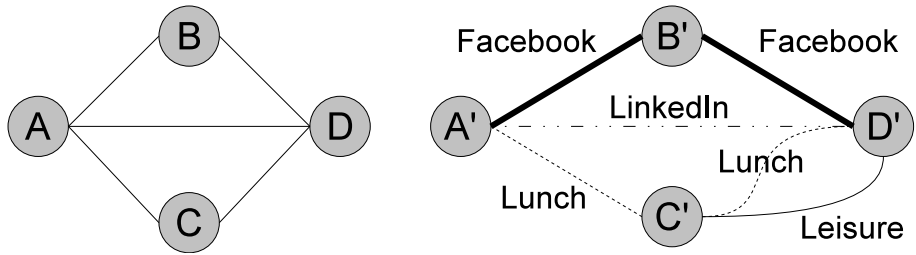}
\caption{A merged multi-layer network (left) and its expanded view revealing different layers}
\label{f:centrality}
\end{figure}

\noindent\textbf{H1} \emph{The centrality of an individual is a function of the different kinds of relationships s/he has with other individuals.} For example, consider Figure~\ref{f:centrality}. Looking at the multi-layer network to the left, the distance between A and D is one because they are directly connected. However, if we consider all the layers hidden behind the multi-layer network (right hand side) our analysis can become much more accurate. First, we can see that the shortest path between A' and D' might not be the direct connection through LinkedIn, but the connection through two Facebook edges, considering that Facebook hosts many more interactions than LinkedIn and thus the two steps on Facebook might be faster to traverse than a single step on LinkedIn. In addition, if we look at node C' we can see that it is in the middle of different possible paths between A' and D', e.g., A' has lunch with C' who goes to the cinema and has lunch with D'. In summary, many hidden paths would contribute to the betweenness of B' and C', depending on the different networks and combinations of networks they are active in.

\noindent\textbf{H2} \emph{Communities can emerge when a specific combination of layers is considered.} Although several community detection algorithms exist, in practice they achieve good results when some more or less well-separated clusters exist. This is strictly related to the way in which community detection algorithms have been defined: some try to maximize modularity, favoring well separated clusters, some use random walk approaches, where the probability that a walker crosses two clusters is proportional to the number of edges between them, some exploit measures like betweenness, that is high when few other edges connect distinct portions of the network \cite{Fortunato2010}. However, when we deal with on-line relationships, community detection becomes extremely hard. According to our hypothesis, this depends on the fact that a large number of semantically different layers are considered all-together (i.e., a super-sociomatrix), determining the co-existence of several hidden communities. 

In Figure~\ref{fig_toyCluster}, the idea is illustrated on a simple example. A network with 8 nodes and 3 types of edges (i.e. a 3-layer network) is depicted in the upper left corner. It is obvious that the analysis of the super-sociomatrix 
does not reveal any interesting patterns as there are too many edges in the graph (in fact, the graph is fully connected in this example). The same can be observed for networks of individual relation types (on the right). However, choosing two specific layers, some more evident clusters emerge (lower left part of the figure, clusters denoted by black and white nodes).

\begin{figure}[h!]
\centering
\includegraphics[width=0.8\textwidth]{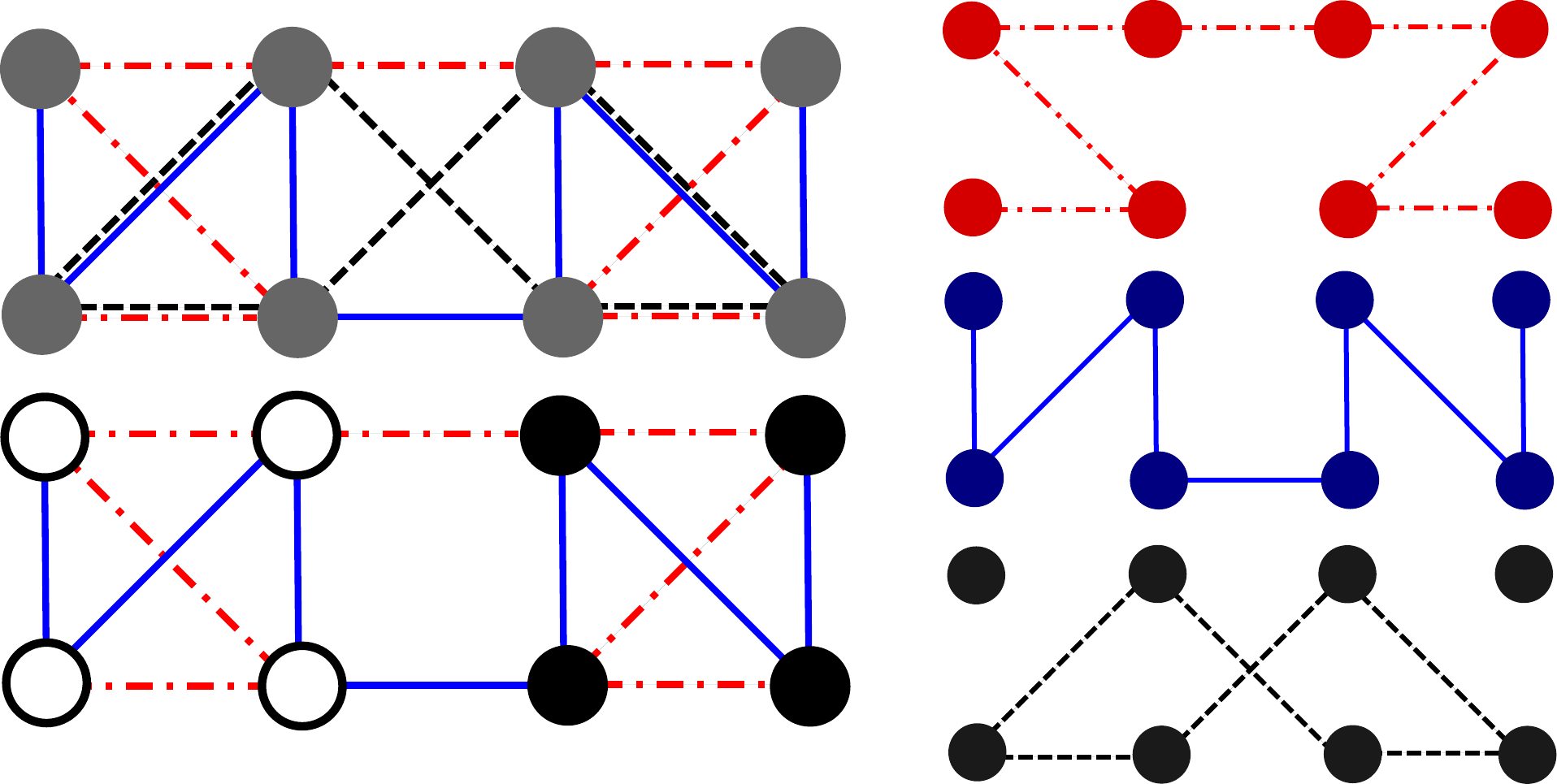}
\caption{Clusterability in combinations of layers. Lower left combination of layers produces an interesting clustering}
\label{fig_toyCluster}
\end{figure}

\noindent\textbf{H3} \emph{Different combinations of layers can be mutually dependent.} As such, being active on two different networks may not have the same importance. For example, we may find that our connections on a specific network are the same people we are connected to on a combination of two other social media sites. As a consequence, staying in this network has a cost in terms of time spent but might not determine any added value with respect to the size of our social experience.

\subsection{Contributions}

In this paper we provide the following contributions:
\begin{itemize}
\item We present a real dataset that depicts five different kinds of relationships between a close community of individuals. The dataset has been collected using both traditional survey-based methods and the analysis of online profiles, and as a consequence it constitutes a multi-layer network spanning both the online and offline dimensions.
\item We apply an existing distance function for multi-layer data \cite{MagnaniSBP13-1} to our dataset. To the best of our knowledge, this is the first time this function is tested on real data.
\item We define a new measure, called \emph{multi-layer betweenness centrality}, that extends betweenness centrality taking into consideration paths crossing several different layers. Also this function is tested on real data, discussing its ability to identify central users with respect to the whole multi-layer structure.
\item We study the variations in our ability to identify meaningful clusters depending on the specific combination of layers under consideration.
\item We relate the concept of network similarity on a power-sociomatrix to the concept of network portfolio, i.e., the management of our stratified social interactions.
\end{itemize}

\subsection{Outline of the paper}

In the next section we introduce the first example of multi-layer network data collected using both online and offline techniques. We present both the adopted methodology and a general description of the data, that will be made publicly available as part of the ICWSM data collection initiative. Then we introduce specific analysis methods and measures taking the combinations of network layers under consideration, including a new betweenness measure for multiple networks. All the measures are experimentally validated on our real data, and we show that while for some measures we obtain a measurable effect, other basic hypotheses cannot be clearly corroborated using the currently available data. This leads to a discussion of future directions concerning both methodological issues and collection of new datasets. We conclude the paper with a review of the state of the art and some concluding remarks.

\section{Dataset}

%

In this section, we are going to introduce a new real-world dataset which consists of measurements of multiple types of relations among the population and can be treated as a multi-layer graph. We first describe the methodology behind data collection, and then provide some basic statistics of the collected data. A full analysis follows, with the application of the new methods introduced in this paper.

\subsection{Methodology}
Collection of data was conducted at the Department of Computer Science at Aarhus University among the employees. The population of the study is 61 employees (out of the total number of 142) who decided to join the survey, including professors, postdoctoral researchers, PhD students and administration staff.

According to \cite{Wasserman1994}, there are two types of variables that can be included in a network data set: \emph{structural} and \emph{composition} variables. Structural variables are measured on pairs of actors and they express specific ties between the pairs (e.g. friendship). Composition variables are defined for individual actors and they are measurements of various actor attributes (e.g. age). 

For our study, we measured 5 structural variables, namely: 
\begin{itemize}
\item current working relationships,  
\item repeated leisure activities,
\item regularly eating lunch together,
\item co-authorship of a publication,
\item and friendship on Facebook.
\end{itemize}
These variables cover different types of relations between the actors based on their \emph{interactions}. All relations are \emph{dichotomous} which means that they are either present or absent, without weights.

Measurements of the first three variables (off-line data) were collected via a \emph{questionnaire} which had been distributed among the employees on-line. The questions were of a \emph{roster} format which means that each actor was presented with a complete list of other actors in the network and asked to select people with whom he/she has the aforementioned ties (working relationship, leisure activities, eating lunch). The number of choices for each question was not limited. The measurements are results of individual assessments done by the actors and therefore the relations are directed, however, we do not intend to study the aspect of personal perception of the relationships and so we decided to flatten the data into nondirectional connections. Thus, if an actor $A$ indicated a tie to actor $B$, we input an edge into the network even if actor $B$ did not indicate a tie to actor $A$.
On the top of this, the respondents were asked to provide their user information for a couple of most widespread online networks. 77\% of the respondents who filled in the questionnaire stated that they have a Facebook account and provided their username. All respondents provided answers to all questions which means that our multi-layer network is complete.

Information about the co-authorship relation was obtained from the on-line DBLP bibliography database without the need to directly ask the actors. A co-authorship of at least one publication by a pair of actors resulted in an edge in the network. DBLP gets new data with a delay of several months and therefore the ``current working relationships" network is quite distinct from it. Moreover, ``current working relationships" network includes other types of interactions than cooperation on papers (e.g. also cooperation on administrative work).

Friendship relations among all the actors who stated that they have a Facebook account were retrieved from the site using a custom application.

\subsection{Data Description}
We describe some common statistical measures of the single-layer networks in Tab.\,\ref{tabDataDescription}. The Co-authorship network is the smallest and less connected of all layers, Work and Lunch networks have the most edges and the highest average vertex degree can be observed for the Facebook network.

\begin{table}[t!] 
\begin{center}
\begin{tabular}{lccccc}
\hline
  &   \small{Work} &  \small{Leisure}  &  \small{Coauthor}  & \small{Lunch} &  \small{FB} \\
\hline
 \small{\# of edges}  &   194  &  88 &  21 &  193 &   124 \\
\small{\# of con. comp.} &  2   &  1   &   8 &   1  &   1    \\
 \small{avg. vertex deg.} &  $6.47$ &   $3.74$  &  $1.68$ &  $6.43$ &   $7.75$   \\ 
\hline
\end{tabular}
\caption{Basic statistics computed on the sociomatrices of the 5 different relations---number of edges, number of connected components and average vertex degree} 
\label{tabDataDescription}
\end{center}
\end{table}

The analysis of the network through a super-sociomatrix approach has been for a long time a viable approach to handle multiple networks. 
In the multi-layer super-sociomatrix, all the layers componing our complex structure of experience are merged together producing a single layer network. We can therefore produce a summary description of this network. It counts 61 nodes and 706 edges belonging to all the 5 layers we are using in this study. Average vertex degree is $11.57$ and the network has a diameter of 4.

\begin{figure}[h!]
\centering
\includegraphics[width=0.9\textwidth]{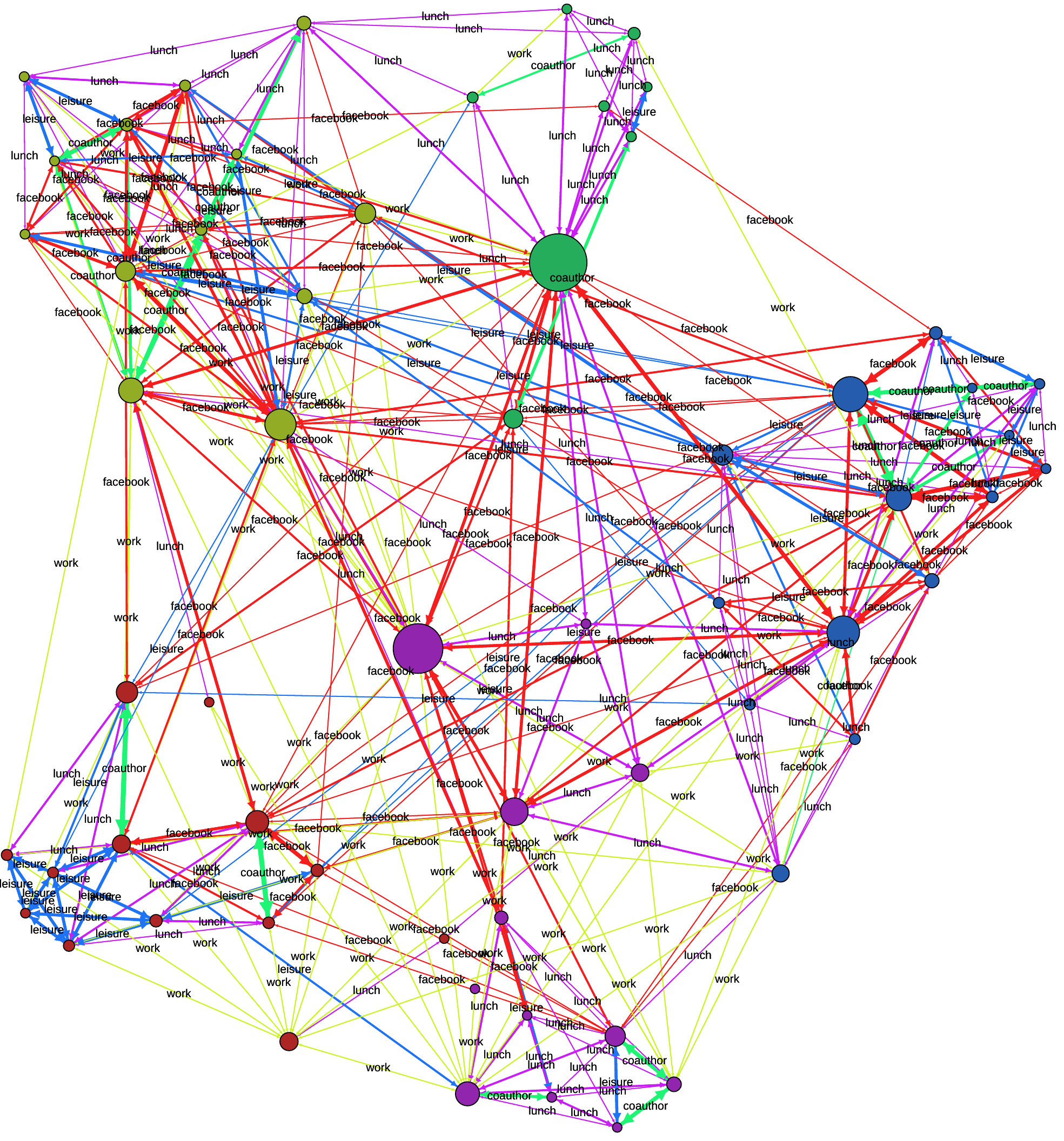}
\caption{Visualization of the super-sociomatrix}
\label{fig_toyCluster}
\end{figure}

\section{Multi-layer Network Analysis}

\subsection{Multi-layer Betweenness}
\begin{figure}[h!]
\centering
\includegraphics[width=.6\columnwidth]{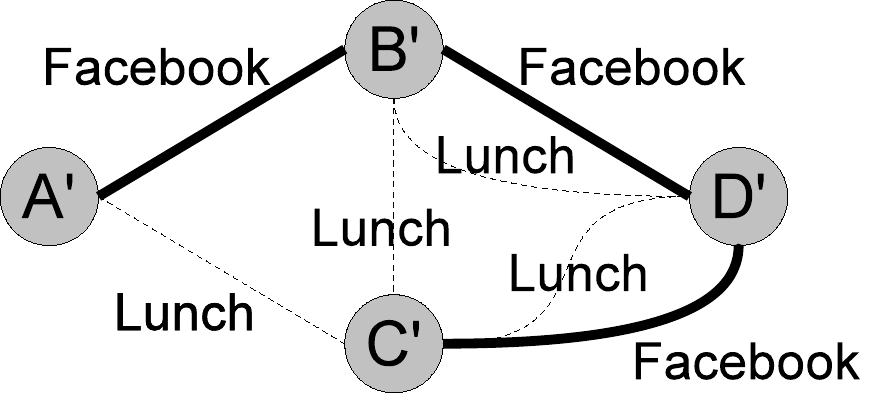}
\caption{Multi-layer distances}
\label{f:pareto}
\end{figure}

Our definition of betweenness is based on the existing concept of geodesic distance for multi-layer networks \cite{MagnaniASONAM2011,MagnaniSBP13-1}.
Therefore, we first introduce this concept by example. Let us consider Figure~\ref{f:pareto}, and the distance between A and D. If we assume to merge the two layers (Facebook and Lunch) and apply a traditional definition of shortest path, we can see that the distance between A and D is 2 (passing through B or through C).

Now, let us consider the two distinct layers. The first consideration is that we have a potential path through B, only traversing Facebook connections, and a potential path through C, only traversing Lunch connections. Therefore, we do not longer say that passing through B or C constitutes two equivalent shortest paths: they represent in fact different and, according to the terminology introduced in \cite{MagnaniSBP13-1}, \emph{incomparable} paths. Then, we can additionally notice that the same sequence of nodes A-B-D can also be traversed through one layer, switch layer and continue. For example, A may send a Facebook message to B, who then has lunch with D and physically delivers the message to him/her.

In summary, the potential shortest paths between A and D on a multi-layer context are:
\begin{itemize}
\item A -lunch$\rightarrow$ C -lunch$\rightarrow$ D
\item A -facebook$\rightarrow$ B -facebook$\rightarrow$ D
\item A -lunch$\rightarrow$ C -facebook$\rightarrow$ D
\item A -facebook$\rightarrow$ B -lunch$\rightarrow$ D
\end{itemize}

Differently from single-network cases, the  \emph{multi-layer distance} between A and D is not defined as a number, but as a set of paths. However, our intuition is that having a way to compute shortest paths in a multi-layer context, we can define an extended version of node betweenness as the number of multi-layer shortest paths between any two nodes that contain the input node.

The first feature of this notion is that it reduces to the existing definition of betweenness when single layers are used. The second feature is that in general we may expect a node to increase its multi-layer betweenness as more arcs are added. Therefore, multi-layer betweenness when compared to single layer betweenness emphasizes if there are nodes that are more or less present in more or less layers. In general, a node with many connections on different layers will have the opportunity of significantly increasing its betweenness thanks to the many alternative paths going through it, while nodes active e.g. on a single layer will more likely keep a low overall value of multi-layer centrality.

\begin{figure}
\centering
\includegraphics[width=\columnwidth]{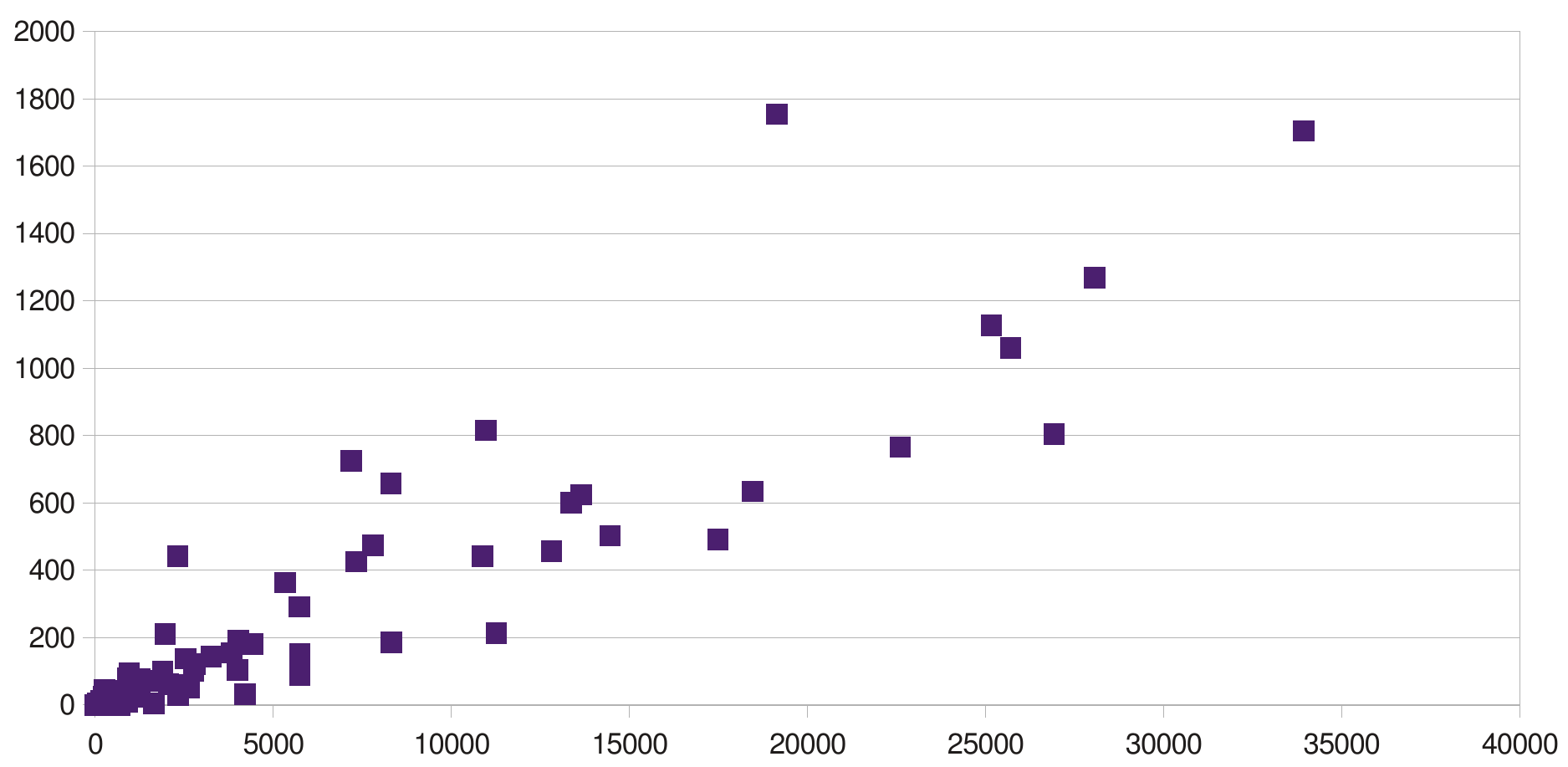}
\caption{Comparison of absolute values of the two different measures of betweenness. Multi-layer betweenness on $x$-axis versus traditional betweenness measure on $y$-axis. Each point corresponds to an actor}
\label{fig_bet_abs}
\end{figure}

Figure~\ref{fig_bet_abs} shows a comparison of traditional betweenness and multi-layer betweenness, for every node in our network. However, given the skewed distribution of these measures a direct comparison of their absolute values is not very informative. Instead, we can compare the ranking of the nodes, e.g., take the node with the highest traditional betweenness and check if it also has one of the highest values of multi-layer betweenness. If the two rankings are very correlated, than we can conclude that our new measure is not significant because it does not highlight any effects of considering all the layers as separate entities.

\begin{figure*}
\centering
\includegraphics[width=\textwidth]{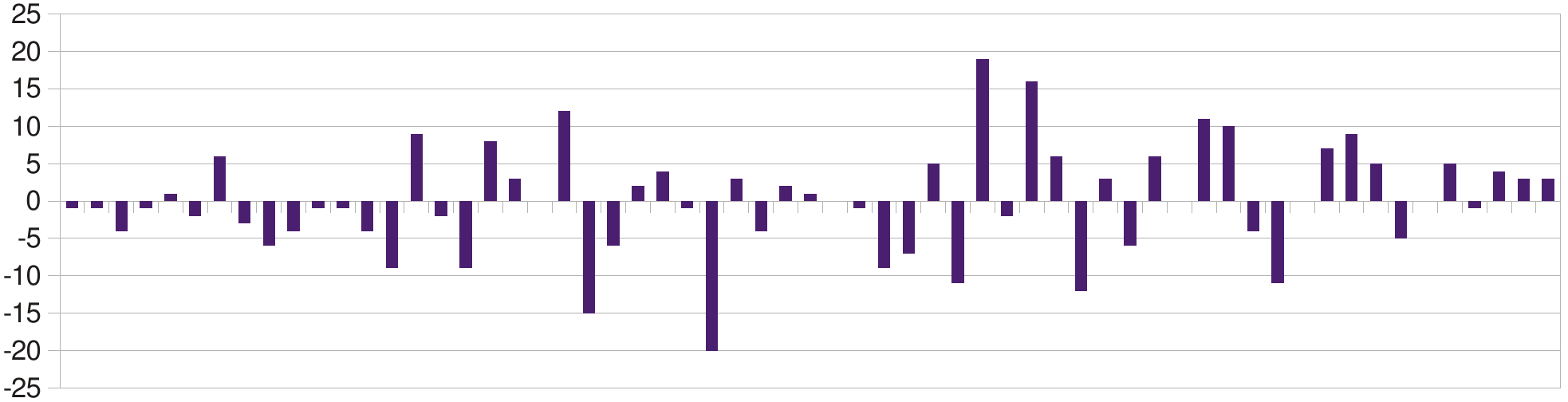}
\caption{Increase/decrease of 61 individual actors in ranking according to two different measures of betweenness. Actors were sorted (from left to right) in a descending order according to the traditional measure of betweenness calculated on the super-sociomatrix. Then, betweenness was recalculated using the new multi-layer distance and increase/decrease in ranking of each actor (compared to the traditional measure) was plotted}
\label{fig_bet}
\end{figure*}

Figure~\ref{fig_bet} shows the changes in ranking using traditional betweenness and multi-layer betweenness. In general, we can observe that these two measures are in fact very correlated. However, there are some cases where the centrality of an individual changes significantly---notice that the sample consists of 61 individuals, and two nodes change their ranking of almost 20 positions, i.e., they completely change their role if the stratified structure is taken into account. At the same time, the "important" nodes remain more or less stable.

\subsection{Multi-layer Clusterability}



\begin{figure}[h!]
\centering
\includegraphics[width=\columnwidth]{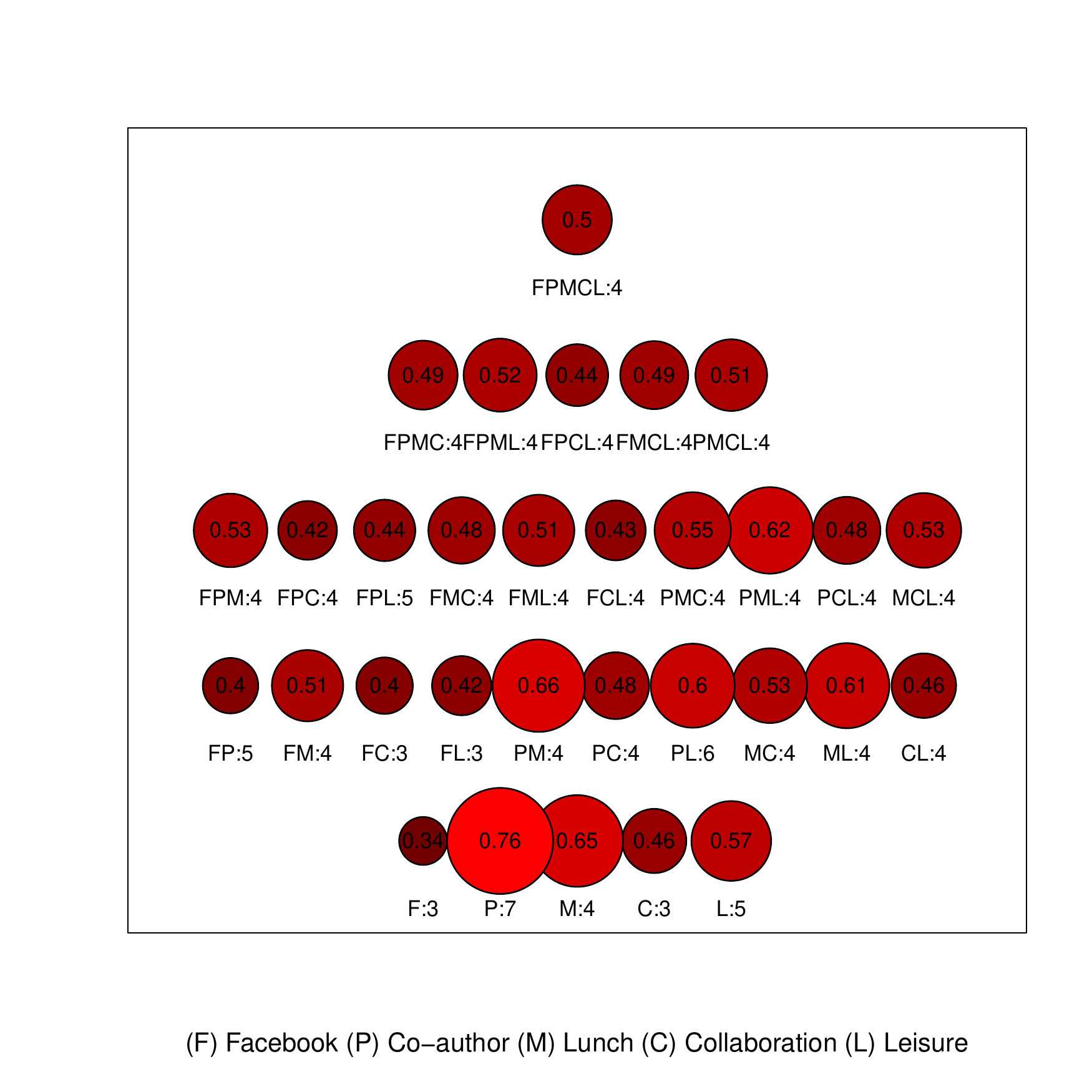}
\caption{Clusterability of all combinations of layers, expressed as the modularity of an identified clustering using a modularity optimization approach}
\label{fig_clusterabilityy}
\end{figure}

At the beginning of the paper we made the hypothesis that communities can emerge when a specific combination of layers is considered. This hypothesis has a potential significant impact on network analysis, because it would identify the source of complexity that makes community detection hard when dealing with large online datasets. Unfortunately, the collected dataset does not highlight clear patterns in this direction.

Figure \ref{fig_clusterabilityy} has been produced as follows. First, we generated all possible combinations of the five layers. Each ball in the figure corresponds to a specific combination, as indicated by the label (e.g., FL corresponds to the combination of (F) Facebook and (L) Leisure networks). Then, for each combination we have computed a clustering based on the optimization of modularity---in particular we have used a multilevel modularity optimization approach. This generated the number of clusters indicated after the label. The higher the modularity, the bigger the ball. The value of modularity is also explicitly indicated inside every ball. For example, the clustering of the co-authorship network (P) has identified 7 clusters, determining a modularity of .76.

On the negative side, we can see that some single layers already achieve a very high level of modularity, showing that they can be easily clustered. This makes it difficult to find clear advantages in exploring hidden combinations of layers.

On the positive side, we can see that the ability of finding clusters decreases when specific layers are added. For example, not considering the collaboration network (C) it is more evident how to cluster the data (FPML, modularity: .52) than when all layers are used (FPMCL , modularity: .50). This difference is small, but looking at Figure~5 we can appreciate how one of the network layers (the one corresponding to the pale color in the figure) introduces some noise on top of an otherwise well identifiable structure with four clusters. As we have mentioned, this effect is however very limited in our dataset, which prevents us from drawing definitive conclusions.

The main counter-hypothesis is that the usage of an offline survey with very well-defined networks and complete answers resulted in a high-quality dataset that does not present clear hidden patterns. To check this hypothesis, the collection of additional sources of online multi-layer data is necessary, and the availability of additional data sources will be a crucial aspect in the development of this research direction.

\subsection{Network Coverage}
It should appear clear from everyone's daily experience that not every new piece of information has the same value. Information is generated out of the possibility to access new data that were unavailable before. Following this very simple concept we assume that new network layers do not necessarily lead to a higher level of information and to a better understanding of underlying phenomena. From this perspective, adding a network layer that duplicates an already existing layer might result in producing no extra information or even in adding redundant and unnecessary data which makes the analysis more complicated. In order to solve this problem, it is therefore necessary to be able to assess the utility of every new layer that is obtained. In other words, it could be measured how much new information a layer brings that is not yet provided in the (combination of) remaining layers.  

We call a perfect combination of layers a perfect \emph{network portfolio}. A perfect network portfolio is a set of layers where every network layer adds a specific and non-redundant information to the construction of the ML model. Obviously, different networks might overlap more or less for many different reasons: the network of collaboration at the department might be a perfect subset of the network of collaboration at the university while the network of the tennis opponents might only partially overlap with the departmental co-workers. In this example, while the departmental network might be replaced by the university network, the tennis network cannot. Therefore, in order to detect the layers that add less information to the ML model, it is important to know to which degree a single network or a combination of networks overlap with another network or a combination of networks. In Tab.\,\ref{tabProb}, we show how different combinations of layers cover all single layers (the best covering cases were selected). Knowing that the combination of Work, Leisure and Facebook layers will cover with a $0.95$ probability the Co-authorship network could lead to the decision not to include the Co-authorship network into our network portfolio.

From a more interpretative perspective, it is interesting to point out that our data show that Facebook is the less coverable network. Even combining all the four remaining networks together \{Work, Leisure, Co-author, Lunch\}, only $0.64$ Facebook edges can be covered by them. This might be a result of the fact that the offline layers are basically goal-oriented networks (groups of people that connect together to reach a specific goal) while Facebook is an online network that can be considered as general-purpose. Connections that have no sense in other layers might exist within Facebook and be, there, perfectly normal. This is why it is difficult to find a combination of offline layers that covers Facebook well.
Results of another measure, Jaccard index of two (multi-layer) networks, are consistent with this finding. Let us define Jaccard index to be the proportion of intersection and union of edges of two networks---thus, it is a measure of similarity. According to this measure, the greatest similarity is between the combinations of layers \{Coauthor, Lunch, Facebook\} and \{Work, Leisure\}---Jaccard index of $0.44$. In Tab.\,\ref{tabJaccard}, for every single layer the most similar combination of other layers can be found. 
The most distinct layer is Co-author which is not surprising since it is a very small network compared to all the other ones. Interestingly, the maximum similarity of Facebook (to a combination of other layers) is only 0.23 (when compared to \{Work, Leisure, Lunch\}), which again implies that it is quite a distinct layer.  

\begin{table}[h!] 
\begin{center}
\begin{tabular}{|l|c|r|}
\hline
 $\mathcal{R}$ & \small{Covering combination of layers} & \small{Prob.} \\
\hline
 Coauthor  & Work, Leisure, FB  & 0.95 \\
 Coauthor  & Work, Leisure & 0.90 \\
 Coauthor  & Work & 0.86 \\
 Leisure &  Work, Coauthor, Lunch, FB   & 0.89 \\
 Leisure &  Work, Lunch   & 0.78 \\
 Lunch &  Work, Leisure, Coauthor, FB & 0.70\\
 Work &  Leisure, Coauthor, Lunch, FB & 0.66\\  
 FB &  Work, Leisure, Coauthor, Lunch & 0.64\\  
\hline
\end{tabular}
\caption{Best covering combinations of layers for each single network. This was computed as conditional probability that there is an edge in the combination of layers in case that there is an edge in the single network} 
\label{tabProb}
\end{center}
\end{table}

\begin{table}[h!] 
\begin{center}
\begin{tabular}{|l|c|r|}
\hline
 $\mathcal{R}$& \small{Similar combination of layers} & \small{Jaccard}\\
\hline
 Lunch &  Work, Leisure & 0.39\\
 Work &  Coauthor, Lunch & 0.36\\  
 Leisure &  Coauthor, Lunch   & 0.27 \\
 FB &  Work, Leisure, Lunch & 0.23\\
 Coauthor  & Leisure, FB  & 0.07 \\  
\hline
\end{tabular}
\caption{Each layer compared to the most similar combination of other layers} 
\label{tabJaccard}
\end{center}
\end{table}


\section{Related work}

Sociological research in on-line communication has always analyzed the multiplicity of identities and this has been one of the classical topics in Computer Mediated Communication studies since the early work of \cite{Turkle1995}. From this perspective, online identity has often been described as a self-conscious performance of identity practices played in different online contexts \cite{boyd2010}. Obviously, this research result, mainly obtained through ethnographic and qualitative approaches, has been largely underestimated when in recent years, researchers have applied SNA techniques to explore large online network data. This has been happening despite the fact that the potential pitfalls of setting up social networks as node-dependent relations have been stressed for a long time \cite{Wasserman1994}.

Nevertheless, the challenge of multiplicity of online experience is still present and beside the simple approaches developed by early researchers \cite{Wasserman1994}, many new and different perspectives to deal with the problem have emerged over the years. Multi-dimensional network approaches \cite{Cai2005,Mucha2010,Berlingerio2011} describe how several relationships between nodes may co-exist, such as friendship, business relationships or shared interests. These relationships are treated as relation networks where every graph represents one type of relation and a network taking into consideration multiple relationships at the same time will be defined as a multi-relational (or heterogeneous) network \cite{Cai2005,Sun2012}. From this perspective, the ability to identify the existing relationships that better help to answer a user's query can also be seen as a problem of feature extraction applied to relational SNA where the hidden relationships have to be exposed \cite{Tang2009}. Some basic data mining approaches have been developed to exploit the multi-relational network structure, specifically clustering (also called \emph{community detection} when nodes represent users) \cite{Yin2006,Sun2009} and link prediction \cite{Leskovec2010,Sun2012}.

Extending this problem to a closer area of application, different types of connections might easily imply a scenario characterized by heterogeneous nodes such as the relationships existing between academics co-authoring papers together or attending the same conferences \cite{DBLP:conf/socialcom/KazienkoBMG10}. These relationships might also be ranked according to several scales, e.g. co-authoring a paper together might be considered more important that attending the same conference. Although heterogeneous networks are not the topic of this paper, the methods and techniques developed to tackle that specific issue might be adapted to multi-layer networks.


\section{Concluding remarks}

In this paper we have claimed that models to represent multiple networks should be complemented with specific analysis tools exploiting their multi-layered structure. So, we have defined some measures aimed at identifying hidden patterns, in particular:
\begin{itemize}
\item a multi-layer version of betweenness centrality, taking paths traversing different layers into consideration,
\item the modularity of all the combinations of layers, to identify hidden communities extending through different layers, and
\item the relative coverage of different combinations of layers.
\end{itemize}
To test these measures, we have performed a data collection initiative building a data set that includes both online and offline networks.

The experimental analysis of our hypotheses has shown that the extended betweenness centrality can indeed identify nodes whose behavior changes significantly when the layered structure of the network is taken into consideration. However, the most important nodes in the network were not affected in the same way. Similarly, the analysis of modularity and relative coverage of combinations of network layers has identified some potentially interesting facts, but still the experimental evidence is too limited to be able to reach a satisfactory conclusion.

In particular, the clustering achievable using a combination of four specific layers is better than the one achievable using the whole super-sociomatrix. However, this does not lead to the identification of different clusters, and the best clusterability can still be found in one of the single layers. In the future, we envision a fine-grained comparison not limited to the values of modularity and to the number of clusters, but focusing on a detailed comparison of node memberships.

These considerations open future directions of research and clearly show that the availability of new real datasets of the kind of the one introduced in this paper are  crucial to progress in this field. In particular, from our analysis it seems that the single layers were already as well defined as to make it difficult to find better patterns only exploiting them partially. On the contrary, the usage of more online data may make it more difficult to identify high quality single layers, giving more importance to the multi-layer based analysis methods introduced in this paper.

As a final remark, all the experiments performed in this work were possible because of the limited size of the data. The scalability of these methods to larger datasets with hundreds of layers, that we may expect to find in online social media, will require optimization efforts that in our opinion will characterize a significant part of the research in this field in the near future.


\end{document}